\documentclass[useAMS,usenatbib,twocolumn]{mn2e}
\usepackage{natbib}
\usepackage{graphicx}
\usepackage{hyperref}
\usepackage{amssymb,latexsym,amsopn}
{\tiny }\usepackage{epsfig}
\usepackage[T1]{fontenc}

\newcommand{\notag}{\nonumber}

\usepackage{color}

\title[]{
Pulsar timing noise and the minimum observation time to detect gravitational waves with pulsar timing arrays
}
\author[Lasky, Melatos, Ravi \& Hobbs]
{Paul D. Lasky$^{1,2}$\thanks{E-mail: paul.lasky@monash.edu}, 
Andrew Melatos$^{2}$, 
Vikram Ravi$^{2,3}$, 
George Hobbs$^{3}$.
\\
$^{1}$Monash Centre for Astrophysics, School of Physics and Astronomy, Monash University, VIC 3800, Australia\\
$^{2}$School of Physics, University of Melbourne, Parkville, Victoria 3010, Australia\\
$^{3}$CSIRO Astronomy and Space Science, Australia Telescope National Facility, PO Box 76, Epping, NSW 1710, Australia
}

\begin{document}

\label{firstpage}

\maketitle

\begin{abstract}
	The sensitivity of pulsar timing arrays to gravitational waves is, at some level, limited by timing noise.
 	Red timing noise -- the stochastic wandering of pulse arrival times with a red spectrum -- is prevalent in slow-spinning pulsars and has been identified in many millisecond pulsars.
	Phenomenological models of timing noise, such as from superfluid turbulence, suggest that the timing noise spectrum plateaus below some critical frequency, $f_c$, potentially aiding the hunt for gravitational waves.  
	We examine this effect for individual pulsars by calculating minimum observation times, $T_{\rm min}(f_{c})$, over which the gravitational wave signal becomes larger than the timing noise plateau.
	We do this in two ways: 1) in a model-independent manner, and 2) by using the superfluid turbulence model for timing noise as an example to illustrate how neutron star parameters can be constrained.
	We show that the superfluid turbulence model can reproduce the data qualitatively from a number of pulsars observed as part of the Parkes Pulsar Timing Array.
	We further show how a value of $f_c$, derived either through observations or theory, can be related to $T_{\rm min}$.  This provides a diagnostic whereby the usefulness of timing array pulsars for gravitational-wave detection can be quantified. 
\end{abstract}

\begin{keywords}
	gravitational waves -- pulsars: general --  stars: neutron -- stars: rotation
\end{keywords}

\section{Introduction}\label{intro}
Pulsar Timing Arrays  \cite[PTAs; e.g.,][]{manchester13,kramer13,mclaughlin13} seek to detect nanohertz gravitational waves from cosmological and extragalactic sources by looking for correlations between contemporaneously measured pulse arrival times from multiple radio pulsars \citep{hellings83}.  The sensitivity of a PTA is limited by pulsar timing noise, i.e., stochastic wandering of pulse arrival times.
External noise sources include interstellar plasma turbulence, jitter noise and errors in terrestrial time standards; see \citet{cordes13} for a description of all dominant noise sources and an estimate of their magnitudes.  Intrinsic noise sources have been attributed to microglitches \citep{cordes85,dalessandro95,melatos08}, post-glitch recovery \citep{johnston99}, magnetospheric state switching \citep[e.g.,][]{kramer06,lyne10}, fluctuations in the spin-down torque \citep{cheng87a,cheng87b,urama06}, variable coupling between the crust and core or pinned and corotating regions \citep{alpar86,jones90}, asteroid belts \citep{shannon13b} and superfluid turbulence \citep{greenstein70,link12a,melatos14}.  Analyses of long-term millisecond pulsar timing data indicate that timing noise power spectra are typically white above some frequency and red below it \citep{kaspi94, shannon10, vanhaasteren11, shannon13a}.

Red timing noise power spectra cannot extend to arbitragerily low frequencies, as the infinite integrated noise-power implies divergent phase residuals and hence (if phase residuals arise from torque fluctuations) unphysical pulsar angular velocities.  One therefore expects the spectrum to plateau, or even become blue, below some turn-over frequency $f_c$.  A number of physical models naturally predict low-frequency plateaus, including superfluid turbulence \citep{melatos14} and asteroid belts \citep{shannon13b}.  We discuss the former in detail below.  A low-frequency plateau enhances prospects for the detection of a stochastic gravitational wave background.  As the gravitational wave spectrum is a steep power law for most cosmological sources \citep[e.g.,][]{maggiore00,phinney01,grishchuk05}, it rises above the plateau below some frequency as long as it too does not have a low-frequency cut-off \citep[e.g.,][and discussion below]{sesana13b,ravi14a}.

In this article, we quantify how a low-frequency timing noise plateau affects the direct detection of gravitational waves with PTAs.  Specifically, we calculate the minimum observation time for any individual pulsar to become sensitive to gravitational wave stochastic backgrounds from binary supermassive black holes (SMBHs) and cosmic strings.
We note this minimum observation time is only an indicative quantity for determining when a gravitational wave signal will dominate the timing residuals for an individual pulsar; it does not account for algorithms that correlate noise properties between pulsars, a point we discuss in more detail throughout.
We do this in two ways, firstly by parametrising the timing noise in a model independent way, and secondly by applying the superfluid turbulence model of \citet{melatos14}.  In the first approach, we express this minimum time in terms of three pulsar observables; the amplitude and spectral index of the timing noise power spectral density and the turn-over frequency.  The second approach is included as an example of how to relate PTA observables to neutron star internal properties in the context of one particular physical model with only two free parameters.  It does not imply any theoretical preference for the superfluid turbulence model, and will be extended to other physical models in the future.   

The paper is set out as follows.  In section \ref{TN} we define a phenomenological model for timing noise, and review predictions for the power spectral density of the phase residuals induced by gravitational waves from SMBHs and cosmic strings.  In section \ref{mintime} we calculate minimum observation times for hypothetical pulsars as a function of their timing noise spectral index, normalisation, and turn-over frequency.  In section \ref{superfluidmodel} we apply the superfluid turbulence model to data and extract `by-eye' parameter estimates for various pulsars in the Parkes Pulsar Timing Array (PPTA).  We then determine criteria for selecting `optimal' pulsars in section \ref{optimal} and conclude in section \ref{conclusion}.

\section{Power spectrum of the Phase residuals}\label{TN}
\subsection{Timing noise}

Let $\Phi_{\rm TN}(f)$ denote the Fourier transform of the autocorrelation function of the phase residuals, $\delta\phi(t)$, viz.
\begin{equation}
	\Phi_{\rm TN}(f)=\int_{-\infty}^{\infty}d\tau\,{\rm e}^{2\pi i f\tau}\left<\delta\phi(t)\delta\phi(t+\tau)\right>.
\end{equation}
If the timing noise is stationary, $\left<\delta\phi(t)\delta\phi(t+\tau)\right>$ is independent of $t$, as is the mean-square phase residual
\begin{equation}
	\left<\delta\phi(t)^{2}\right>=\frac{1}{\pi}\int^{\infty}_{0}df\,\Phi_{\rm TN}(f)\label{phaseresidual}.
\end{equation}
In practice, the time spent observing the neutron star, $T_{\rm obs}$, is finite.  Hence, one must replace the lower terminal of the integral in the right-hand side of (\ref{phaseresidual}) by $f_{\rm obs}\equiv1/T_{\rm obs}$.  In reality, fitting models to timing data implies PTAs are sensitive to $f<f_{\rm obs}$ [see \citet{coles11} and \citet{vanhaasteren13} for details of timing-model fits in the presence of red noise] implying the lower terminal in (\ref{phaseresidual}) depends on the PTA data analysis algorithm, with $f\lesssim f_{\rm obs}$.  

Millisecond pulsar radio timing experiments measure $\Phi_{\rm TN}(f)\propto f^{-q}$ at low frequencies, $f\lesssim1\,{\rm yr}^{-1}$, with $q\geq0$ \citep[e.g.,][]{kaspi94, shannon10, vanhaasteren11, shannon13a}.  However, the observed power law must roll over below some frequency, $f_{c}$, otherwise equation (\ref{phaseresidual}) implies divergent phase residuals.  To capture this phenomenologically, we model the spectrum in its entirety by
\begin{equation}
	\Phi_{\rm TN}(f)=\frac{A_{\rm TN}}{\left(1+f^{2}/f_{c}^{2}\right)^{q/2}} + A_{\rm W},\label{phi}
\end{equation}
which has the observed large-$f$ behaviour and is even in $f$.  In equation (\ref{phi}), $A_{\rm TN}$ (with units of time) is the dc power spectral density, i.e. $\Phi_{\rm TN}\left(f\ll f_{c}\right)= A_{\rm TN}$, which cannot be measured directly in existing data sets 
\citep{shannon13a}.  In the regime where $\Phi_{\rm TN}(f)\propto f^{-q}$, we can express the more commonly used root-mean-square-induced pulsar timing residuals, $\sigma_R$, in terms of $A_{\rm TN}$ and $q$ as
\begin{equation}
	\sigma_R=
		\frac{5.64}{\sqrt{q-1}}\left(\frac{A_{\rm TN}}{10^{-10}\,{\rm yr}}\right)^{1/2}\left(\frac{f_{\rm obs}}{1\,{\rm yr^{-1}}}\right)^{1/2}\left(\frac{P}{1\,{\rm ms}}\right)\,{\rm ns}.
		\label{convert}
\end{equation}
where $P$ is the pulsar spin period.  For completeness, we include a white noise component, $A_{\rm W}$, in equation (\ref{phi}), which is observed in all pulsars, dominates for $f\gtrsim1\,{\rm yr}^{-1}$, and is the only observed noise component in some objects.  The white component contributes weakly to setting the minimum observation time for gravitational wave detection by PTAs, the key concern of this article.

Equation (\ref{phi}) can be compared against predictions of phase residuals from the cosmological gravitational wave background, $\Phi_{\rm GW}(f)$.  The reciprocal of the frequency where the two curves intersect gives the minimum observation time, $T_{\rm min}$, required  
before an individual pulsar becomes sensitive to a gravitational wave background,
\begin{equation}
	\Phi_{\rm TN}\left(T_{\rm min}^{-1}\right)=\Phi_{\rm GW}\left(T_{\rm min}^{-1}\right).\label{Tmin}
\end{equation}

Equation (\ref{Tmin}) provides a quantitative method for determining when the gravitational wave signal will dominate the timing residual power spectrum.  We emphasise that this is only an indicative threshold for detection; it is not a substitute for a careful signal-to-noise estimate given desired false alarm and false dismissal rates.  Cross-correlation search algorithms look simultaneously at a range in $f$ \citep[e.g.,][]{hellings83,jenet05,anholm09,vanhaasteren09}.  For example, our definition (\ref{Tmin}) is equivalent to the boundary between the `weak signal limit' and the `intermediate regime' as defined in \citet{siemens13}.
While \citet{siemens13} calculate a scaling of gravitational wave detection significance with time assuming only white timing noise, they also perform simulations with red noise assuming $q=-3$.  
A future research project is therefore to introduce red noise with and without a low-frequency turn-over into the analytic calculations of \citet{siemens13}.  

It is likely that the near future will see an increasing number of PTA pulsars satisfy the condition $\Phi_{\rm GW}(f)>\Phi_{\rm TN}(f)$, and that this will occur {\it before} a statistically significant detection is announced.  Equation (\ref{Tmin}) and the analysis presented in this article therefore provide an important input into the time-scale on which this condition will be met by individual pulsars, as a prelude to a cross-correlation detection strategy.

\subsection{Cosmological gravitational wave background}\label{SGWB}
Pulsar timing arrays are sensitive to gravitational wave backgrounds generated by two cosmological sources\footnote{Relic gravitational waves from inflation, such as those purportedly seen by the BICEP2 experiment \citep{ade14}, are expected to be undetectably weak in the pulsar timing band, but may be relevant for Advanced LIGO; see \citet{aasi14} and references therein.}: binary supermassive black holes and vibrations from cosmic strings.

\subsubsection{Supermassive binary black holes}
At binary separations where gravitational radiation dominates the orbital dynamics, the SMBH background is parametrised as a power law
\begin{equation}
	h_{c}(f)=A_{\rm GW}\left(\frac{f}{\mbox{yr}^{-1}}\right)^{\alpha},\label{hc}
\end{equation}
with $\alpha=-2/3$ \citep{phinney01}.  The normalisation coefficient, $A_{\rm GW}$, is the subject of intense debate.  We utilise the most recent predictions by \citet{sesana13} and \citet{ravi15}, quoted in table \ref{table}.  These two predictions assume that gravitational wave emission has already circularized the binary orbits; at binary separations where energy loss to environments dominates instead, the SMBH wave-strain spectrum whitens \citep{sesana13b,ravi14a}.  Whitening of $\Phi_{\rm GW}(f)$ at low frequencies increases $T_{\rm min}$.  

The one-sided power spectral density of the pulsar phase residuals induced by $h_{c}(f)$ is given by
\begin{equation}
	\Phi_{\rm GW}(f)=\frac{h_{c}(f)^{2}}{12\pi^{2}P^{2}f^{3}}. \label{PhiGW}
\end{equation}
$\Phi_{\rm GW}(f)$ has units of time and can be compared directly with $\Phi_{\rm TN}(f)$ as in equation (\ref{Tmin}).

\subsubsection{Cosmic strings}
Cosmic strings are topological defects that may form in phase transitions in the early Universe and produce strong bursts of gravitational radiation, which may be detectable in PTAs \citep{damour00,damour01,damour05}.  A cosmic string-induced stochastic background of gravitational waves is characterised by three dimensionless parameters: the string tension, $G\mu$, the reconnection probability, $p$, and a parameter, $\epsilon$, related to the size of loops.  The best quoted limit of $G\mu\lesssim1.2\times10^{-8}$ is derived from PTA limits of the stochastic gravitational wave background \citep{vanhaasteren11,vanhaasteren12}, although a more stringent constraint (still to be computed), is possible with existing data sets [see \citet{sanidas13} for projected constraints in the near future].  Combined observations using the ground-based Laser Interferometer Gravitational Wave Observatory (LIGO) and Virgo constrain the $\epsilon$ -- $G\mu$ plane to be $7\times10^{-9}<G\mu<1.5\times10^{-7}$ and $\epsilon<8\times10^{-11}$ \citep{abbott09,aasi14}.  Limits on $G\mu$ are model dependent; the reconnection probability is inversely proportional to $\Phi_{\rm GW}(f)$, and smaller values of $\epsilon$ increase the minimum gravitational wave frequency emitted.  This can take the maximum of the stochastic background out of the sensitivity band for PTAs \citep[e.g.,][]{siemens07,olmez10}.  

Despite the above caveats, a power-law model for the characteristic strain spectrum from cosmic strings given by equation (\ref{hc}) with $-1\lesssim\alpha\lesssim0.8$ is a good approximation for the PTA frequency band \citep[][]{maggiore00}.  The predicted range for $A_{\rm GW}$ is quoted in Table \ref{table}.

\begin{table*}
\centering
\begin{tabular}{|cccc|}
	\hline\hline
	Source & $A_{\rm GW}^{\rm min}$ & $A_{\rm GW}^{\rm max}$ & $\alpha$\\
	\hline
	SMBHs \citep[][68\%]{sesana13} & $3.5\times10^{-16}$ & $1.5\times10^{-15}$ & $-2/3$\\
	SMBHs \citep[][95\%]{ravi15} & $5.1\times10^{-16}$ & $2.4\times10^{-15}$ & $-2/3$\\
	Cosmic Strings & $10^{-16}$ & $10^{-15}$ & $-1\lesssim\alpha\lesssim-0.8$\\
	\hline
\end{tabular}
\caption{\label{table} 
	Theoretical spectral parameters for the gravitational wave background from supermassive black hole binaries (SMBHs) and cosmic strings.  Three predictions for the SMBH population are presented: the 68\% confidence interval from \citet{sesana13}, and the 95\% confidence interval 
	from \citet{ravi15}.  The cosmic string models are from \citet{maggiore00}.
	}
\end{table*}

\section{Model Independent Minimum Observation Time}\label{mintime}
To attain adequate sensitivity to gravitational waves at a frequency, $f$, in the phase residuals of an individual pulsar, we must have $\Phi_{\rm GW}(f)>\Phi_{\rm TN}(f)$ for that pulsar, subject to the caveats regarding specific data analysis algorithms expressed in the text following equation (\ref{Tmin}).  If equation (\ref{hc}) applies across all relevant frequencies, and $\Phi_{\rm TN}(f)$ turns over below $f_{c}$, then $\Phi_{\rm GW}(f)>\Phi_{\rm TN}(f)$ is always satisfied for some $f=T_{\rm min}^{-1}$, as in equation (\ref{Tmin}).  

\begin{figure*}
	\includegraphics[angle=0,width=0.95\linewidth]{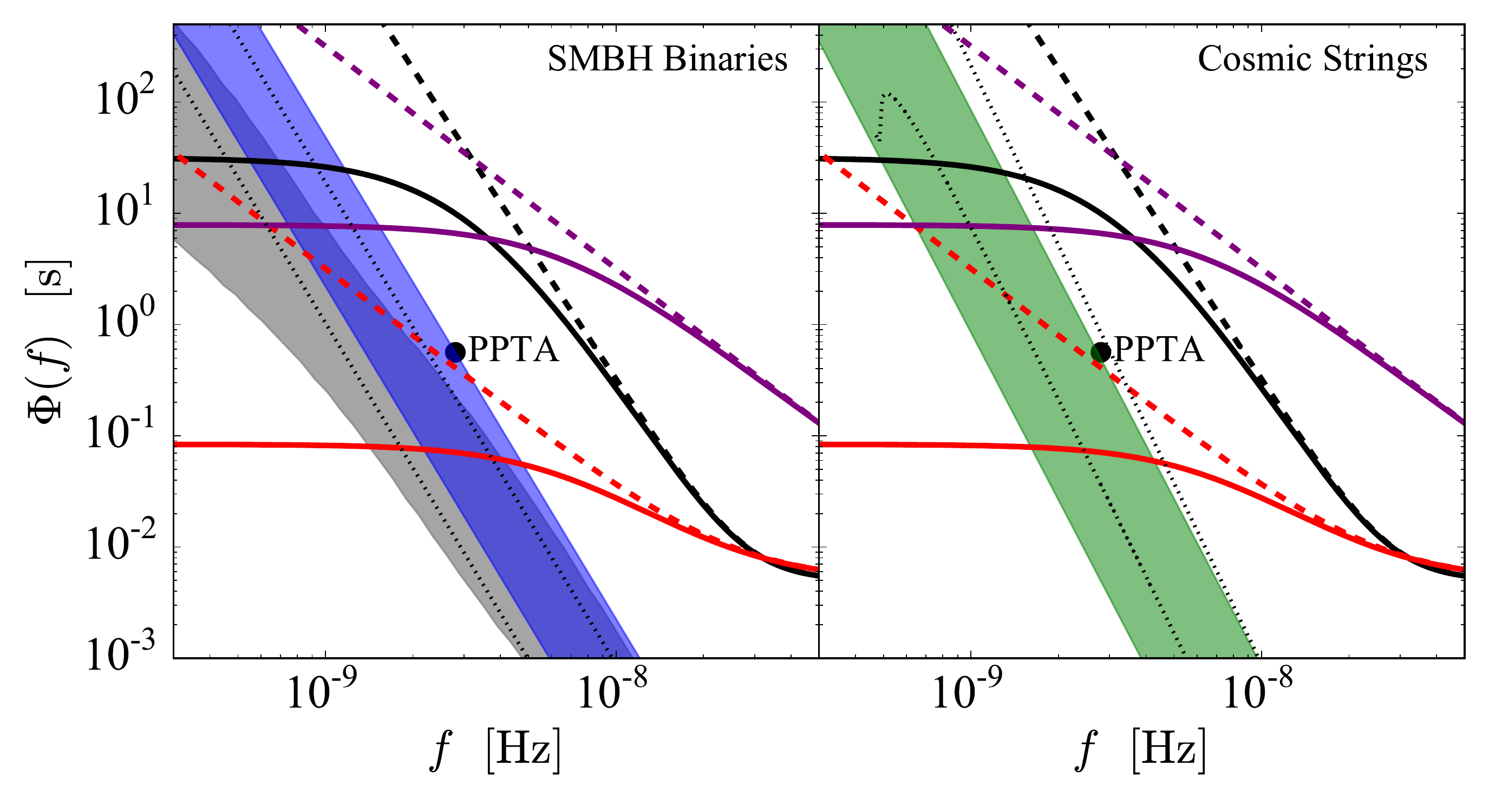}
	\caption{\label{phifig} Power spectral density of timing noise phase residuals, $\Phi_{\rm TN}$, and gravitational wave phase residuals, $\Phi_{\rm GW}$ (both in s) as functions of frequency, $f$ (in Hz).  In both plots, the black, red and purple curves are indicative examples of pulsars described by equation (\ref{phi}) with $(f_{c},\,A_{\rm TN},\,q)=(0.1 {\rm yr}^{-1},\,10^{-10}\,{\rm yr},\,4)$, $(0.2 {\rm yr}^{-1},\,10^{-10}\,{\rm yr},\,2)$ and $(0.2 {\rm yr}^{-1},\,10^{-8}\,{\rm yr},\,2)$ respectively and with a nominal white noise component of $A_{\rm W}=10^{-5}$\,s.  The solid curves include the turnover at $f_c$ in equation (\ref{phi}), and the dashed curves extrapolate backwards the high-$f$ scaling $\propto f^{-q}$ (i.e., $f_{c}\rightarrow0$).   The shaded regions and the dotted black curves encompass the regions covered by $\Phi_{\rm GW}$ for the parameter ranges quoted in Table \ref{table}, where the left-hand plot is for SMBHs and the right plot is for cosmic strings.  In the left plot, the shaded blue region 
	is the 95\% confidence interval from \citet{ravi15}, the region enclosed by the black dotted curves is the 68\% confidence interval from \citet{sesana13} and the shaded grey region is the predicted range from \citet{ravi14a}.  In the right plot, the shaded green region represents cosmic string models with $10^{-16}<A_{\rm GW}<10^{-15}$ and $\alpha=-1$.  The dotted black curves have $p=\epsilon=1$ and $G\mu=1.2\times10^{-8}$ and $1.0\times10^{-10}$ for the top and bottom curves respectively.  The black dot labelled `PPTA' is the lowest published limit on the stochastic background \citep{shannon13a}. 
	}
\end{figure*}

In figure \ref{phifig} we plot $\Phi_{\rm TN}$ and $\Phi_{\rm GW}$ as functions of $f$.  The coloured shaded regions and the region enclosed by the black dotted curves in the left-hand plot contain all the $\Phi_{\rm GW}$ curves in the parameter range in Table \ref{table} for SMBHs.  The blue shaded region 
is the 95\% confidence interval 
from \citet{ravi15} as described in table \ref{table}.  The black dotted curves enclose the 68\% confidence interval from \citet{sesana13}.  The shaded grey region is the predicted range from \citet{ravi14a} that includes low-frequency-whitening of $\Phi_{\rm GW}(f)$ due to non-circular binaries.  In the right-hand plot, the green shaded region is the parameter space enclosed by the cosmic string predictions from table \ref{table} with $\alpha=-1$.  The dotted black curves are specific, representative calculations of the cosmic string background with $p=\epsilon=1$ and $G\mu=1.2\times10^{-8}$ (top curve) and $1.0\times10^{-10}$ (bottom curve)\footnote{Calculations used the GWPlotter website: \hyperref[http://homepages.spa.umn.edu]{http://homepages.spa.umn.edu/$\sim$gwplotter} }.  The black, red and purple curves in each panel are indicative examples of pulsar timing noise as described by equation (\ref{phi}), with values of $f_{c}$, $A_{\rm TN}$, $q$ and $A_{\rm W}$ given in the caption to figure \ref{phifig}.  The correspondingly coloured dashed curves extrapolate backwards the power-law scaling (equivalently assuming $f_{c}\rightarrow0$).  Finally, the dot labelled `PPTA' in both panels marks the lowest limit on the stochastic gravitational wave background from \citet{shannon13a}.

Figure \ref{phifig} illustrates the principal idea of this paper.  If $\Phi_{\rm TN}$ is a simple power-law without a low-frequency turn-over, and for moderate values of $q$, timing noise masks the gravitational wave background down to low frequencies, and $T_{\rm min}$ is correspondingly long (we quantify this below).  A turn-over in $\Phi_{\rm GW}(f)$ at some $f_{c}$ is therefore critical for practical PTA experiments with any millisecond pulsar that exhibits a steep timing noise spectrum with $q\gtrsim2$.  The low-frequency plateau in $\Phi_{\rm GW}$ from elliptical binary SMBHs (the grey shaded region in the left-hand panel of figure \ref{phifig}) makes the need for a turn-over in $\Phi_{\rm TN}(f)$ even more acute.

\begin{figure*}
	\includegraphics[angle=0,width=0.95\linewidth]{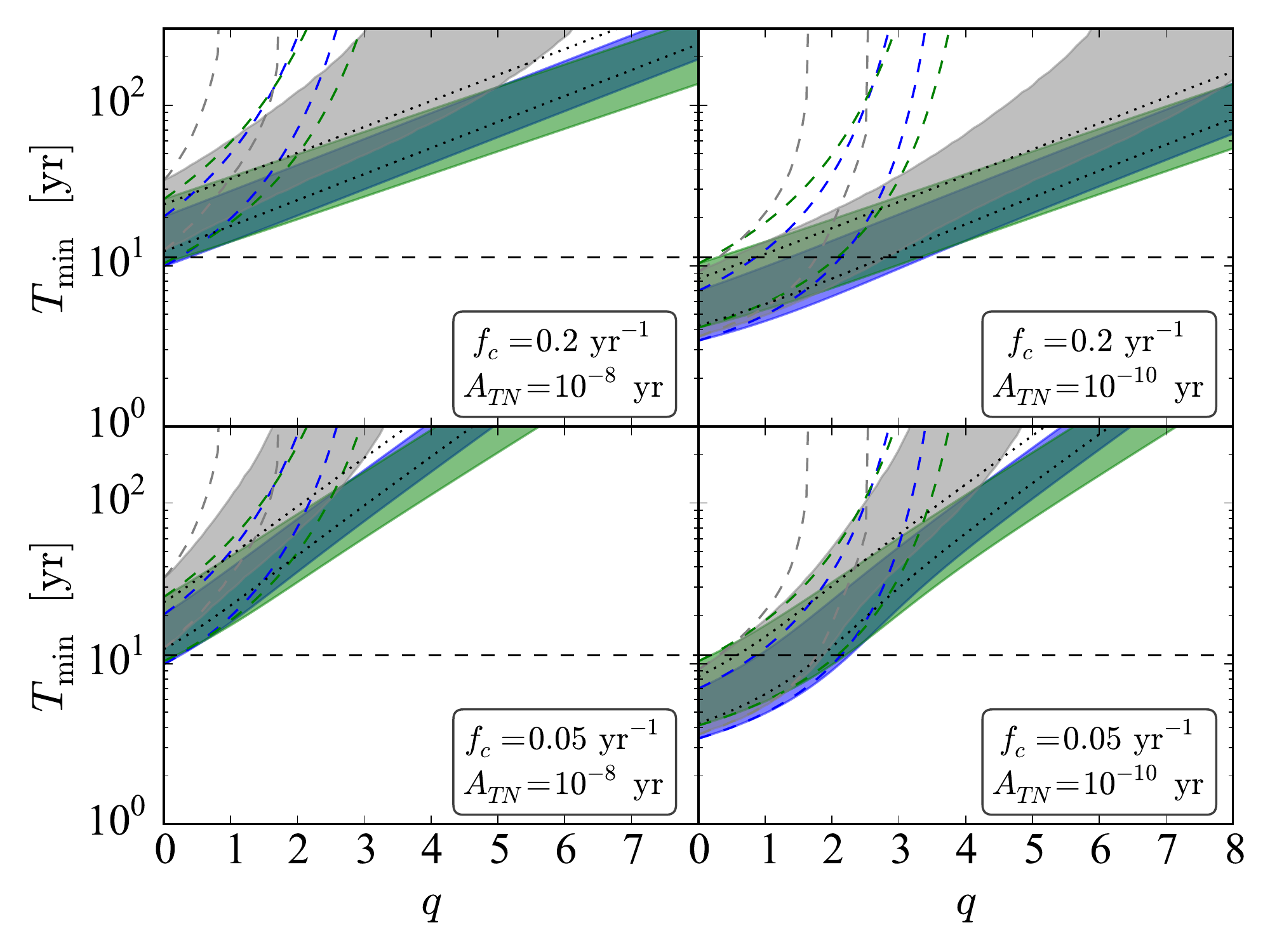}
	\caption{\label{ObsTime} Minimum observation time, $T_{\rm min}$, defined by equation (\ref{Tmin}), as a function of  the high-frequency timing noise spectral index, $q$, with different values of $f_{c}$ and $A_{\rm TN}$ in each panel.  The shaded blue and grey regions represent the ranges of solution space for $T_{\rm min}(q)$ for binary SMBHs from \citet{ravi15} and \citet{ravi14a} respectively, while the dotted black curves encompass the solution space predicted by \citet{sesana13}.
	The shaded green regions are cosmic string predictions.  The coloured dashed curves are the corresponding limits where the timing noise spectrum does not plateau at $f\lesssim f_c$.  The horizontal dashed black line is longest the PPTA observing time in published data, viz. $11.3$ yr \citep{shannon13a}.}
\end{figure*}

In figure \ref{ObsTime} we plot the minimum observation time, $T_{\rm min}$, defined by equation (\ref{Tmin}), as a function of the asymptotic (high-$f$) timing noise spectral index, $q$, for a hypothetical pulsar with $P=10\,{\rm ms}$ and various values of $f_{c}$ and $A_{\rm TN}$ in each panel.  The shaded regions
and dotted black curves delineate the ranges of $T_{\rm min}(q)$ for binary SMBHs and cosmic strings, following the same colour scheme as in figure \ref{phifig} and as detailed in the caption of figure \ref{ObsTime}.  The coloured dashed curves give the limits on $T_{\rm min}(q)$ if $\Phi_{\rm TN}$ does not turn-over (i.e., $f_{c}\rightarrow0$).  The horizontal dashed black line marks the PPTA observing time of 11.3\,yr used for the lowest limit on the stochastic background published to date \citep{shannon13a}.

To help interpret figures \ref{phifig} and \ref{ObsTime}, consider a hypothetical pulsar with $A_{\rm TN}=10^{-8}$ yr (i.e., the two left hand panels) and $q=2$.  If the timing noise spectral density turns over at $f_{c}=1/5$ yr$^{-1}$ or $1/20$ yr$^{-1}$, the minimum observation time given the most optimistic scenario from \citet{ravi15} is $T_{\rm min}=20.5$ yr or $37.2$ yr, respectively.  On the other hand, if $\Phi_{\rm TN}(f)$ does {\it not} turn over, then the dashed blue curves show that the pulsar 
is insensitive to a gravitational wave signal
until $T_{\rm min}=70.7$ yr.  The effect of a plateau in $\Phi_{\rm TN}(f)$ is therefore quite striking.  Pulsars without a plateau and $q\gtrsim3$ (depending less sensitively on $A_{\rm TN}$) are relatively inferior as a tool for detecting gravitational waves.

\section{Timing Noise from Superfluid Turbulence: A Worked Example}\label{superfluidmodel}
In section \ref{mintime}, the description of timing noise is model independent, in the sense that $\Phi_{\rm TN}$ is parametrised phenomenologically by equation (\ref{phi}), without reference to a specific underlying, physical model.  In this section, we repeat the analysis in section \ref{mintime} for the timing noise model of \citet{melatos14} and \citet{melatos14b}, which attributes the fluctuating phase residuals to shear-driven turbulence in the interior of the neutron star.  We emphasise that we do not express any theoretical preference for this model ahead of other models in the literature (see section \ref{intro}).  We focus on it here only because (i) it is predictive, (ii) its results can be expressed in compact, analytic form and, (iii) the theoretical formula for $\Phi_{\rm TN}(f)$ depends on just three internal neutron star parameters, so it is easy to infer constraints on these parameters by combining the model with data.

Consider an idealised neutron star model in which the rigid crust is coupled to the charged electron-proton fluid which, in turn, couples through mutual friction to the inviscid neutron condensate.  The electromagnetic braking torque creates a crust-core shear layer that excites turbulence in the high-Reynolds number superfluid \citep{peralta05,peralta06a,peralta06b,melatos07,peralta08}.  The turbulent condensate reacts back to produce angular momentum fluctuations in the crust, which are observed as timing noise \citep{greenstein70,melatos10}.  In particular, \citet{melatos14b} showed that the timing noise spectral density can be expressed as 
\begin{eqnarray}
	\Phi_{\rm TN}\left(f\right)\!\!\!\!\!\!&=\!\!\!\!\!\!&\frac{15\Gamma(q/2)}{8\pi^{1/2}\Gamma\left[(q-1)/2\right]\lambda^{2}\eta\left(R^{-1}\right)}\notag\\
		\!\!\!\!\!\!&\times\!\!\!\!\!\!&\int^{\infty}_{2\pi}dx\, x^{-35/3}\left(x^{4}+3x^{2}+9\right)\left[1+\frac{4\pi^{2}f^{2}}{\eta\left(R^{-1}\right)^{2}x^{4/3}}\right]^{-q/2},\label{superfluidTN}
\end{eqnarray}
where $\Gamma(x)$ is the Gamma function.  Equation (\ref{superfluidTN}) contains three free parameters: the non-condensate fraction of the moment of inertia, $\lambda=I_{c}/I_{0}$, the decorrelation time-scale, $\eta(R^{-1})^{-1}$, and $q$.  Here, $I_{c}$ is the moment of inertia of the crust plus the rigidly rotating charged fluid plus entrained neutrons, $I_{0}$ is the total moment of inertia, and we define $\eta(R^{-1})=(2\pi)^{-1/2}\epsilon^{1/3}R^{-2/3}\gamma$, where $\epsilon$ is the energy dissipation rate per unit enthalpy (which, in general, is a function of the spin-down rate), $\gamma=\tau_{\rm eddy}/\tau_{\rm turb}\le1$ is the ratio of the eddy turnover time-scale to the characteristic time-scale over which turbulent structures change (which is longer in general due to pinning), and $R$ is the stellar radius.   

The value of the exponent, $q$, in equation (\ref{superfluidTN}) depends on the form of the superfluid velocity two-point decorrelation function.  \citet{melatos10} executed a first attempt to calculate the velocity correlation function numerically on the basis of Hall-Vinen-Bekarevich-Khalatnikov superfluid simulations \citep{peralta08}, but it is not well understood for terrestrial turbulence experiments, let alone for a neutron star interior, especially when stratification plays a role \citep[e.g.,][and references therein]{lasky13a}.  An empirical choice is therefore made that reproduces the asymptotic power-law dependence from timing noise data, i.e., $\Phi_{\rm TN}\propto f^{-q}$ as $f\rightarrow\infty$ [for details see \citet{melatos14,melatos14b}].  We emphasise equation (\ref{superfluidTN}) is not a unique choice, nor can it be inverted uniquely to infer the underlying velocity correlation function \citep{melatos14b}.

In addition to the power-law scaling at high-frequencies, the superfluid turbulence model predicts a plateau at $f\lesssim f_{c}\approx\eta(R^{-1})$.  For time intervals greater than $\sim1/\eta(R^{-1})$, turbulent motions throughout the star decohere, implying torque fluctuations exerted on the crust become statistically independent.  By expanding equation (\ref{superfluidTN}) for $f\ll f_{c}$ and $f\gg f_{c}$, and evaluating the resultant expression in terms of equation (\ref{phi}), we find
\begin{eqnarray}
A_{\rm TN}\!\!\!\!\!\!&=&\!\!\!\!\!\!\frac{9\Gamma(q/2)}{16(2\pi)^{67/6}\sqrt{2}\Gamma[(q-1)/2]\eta(R^{-1})\lambda^{2}}\notag\\
&&\times\left(16\pi^{4}+\frac{120\pi^{2}}{13}+\frac{45}{8}\right),\label{superfluidTNapproxATN}\\
f_{c}\!\!\!\!\!\!&=&\!\!\!\!\!\!\frac{\eta(R^{-1})}{\left(2\pi\right)^{1/3}}\Big[-10\left(\frac{16\pi^{4}}{q-10}+\frac{12\pi^{2}}{q-13}+\frac{9}{q-16}\right)\notag\\
	&&\,\,\,\,\,\,\,\,\,\,\,\,\,\,\,\,\,\,\,\,\times\left(16\pi^{4}+\frac{120\pi^{2}}{13}+\frac{45}{8}\right)^{-1}\Big]^{1/q}.\label{superfluidTNapproxfc}
\end{eqnarray}
Equations (\ref{superfluidTNapproxATN}) and (\ref{superfluidTNapproxfc}) relate the phenomenological model in section \ref{mintime} to the specific physical model in this section.  A similar approach applies equally to other models.

\begin{figure}
	\includegraphics[angle=0,width=0.95\columnwidth]{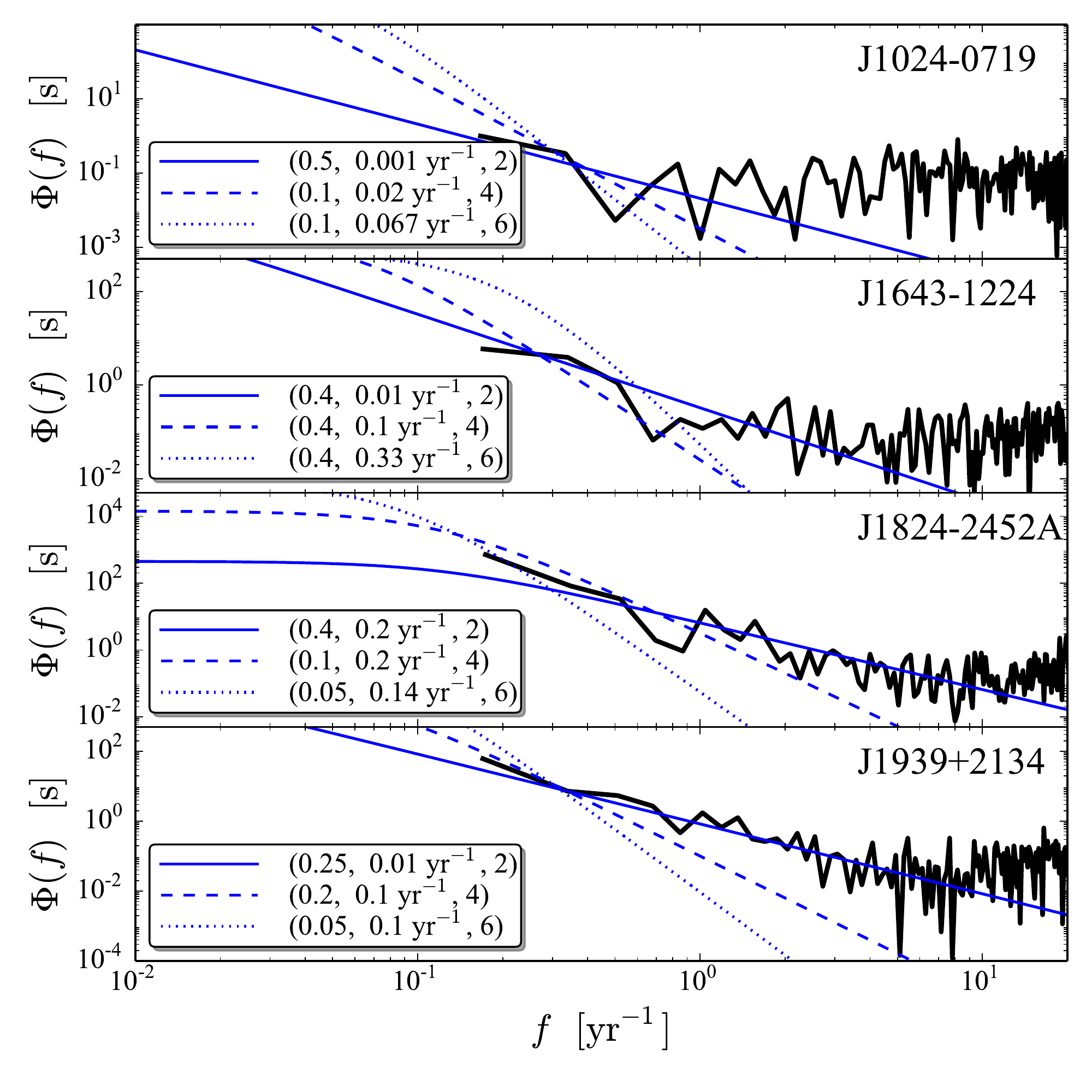}
	\caption{\label{superfluidPSD1} 
	Phase residual power spectra, $\Phi(f)$ for four millisecond pulsars from the Parkes Pulsar Timing Array (thick black curves).  Overplotted are theoretical curves generated by the superfluid turbulence model for the physical parameters  $\left(\lambda,\,\eta,\,q\right)$ specified in the legend. The corresponding values of $A_{\rm TN}$ and $f_{c}$ for each curve are shown in figure \ref{fcAtn}.
	}   
\end{figure}

In figure \ref{superfluidPSD1} we show four examples of millisecond pulsar phase residual power spectra measured by the PPTA \citep{manchester13}.
Overplotted on the data are reasonable `by-eye' fits generated by the superfluid turbulence model for $q=2$, 4 and $6$.  The fits are neither unique nor optimal (e.g., in a least-squares sense), but they are representative.  It is outside the scope of this paper to extract detailed fits and values for $\lambda$, $\eta(R^{-1})$, and $q$ for each pulsar\footnote{The amplitude and spectral index of red-noise in pulsar timing residuals are highly covariant, especially when only the lowest few frequency bins show evidence for red noise \cite[e.g.,][]{vanhaasteren09, vanhaasteren13}.  Finding best-fit parameters for the superfluid turbulence model is therefore a non-trivial task that will be the subject of future work.}.  We simply note that a broad range of parameters fit the phase residuals for any given pulsar.
The pulsars shown in figure \ref{superfluidPSD1} have been chosen as they appear to have moderate to high levels of timing noise, cf. other PPTA pulsars.  All exhibit a relatively red spectrum.  In the context of superfluid turbulence, they imply $f_{c}\gtrsim10^{-2}$ yr$^{-1}$, so that the plateau is potentially observable in the not-too-distant future\footnote{We note that PSR J1824$-$2452A resides in a globular cluster \citep{lyne87}, implying most of the timing noise is likely a result of motions within that cluster rather than superfluid turbulence.  The curves shown in figure \ref{superfluidPSD1} therefore represent an upper limit on the contribution from superfluid turbulence.}.  

\begin{figure}
	\includegraphics[angle=0,width=0.95\columnwidth]{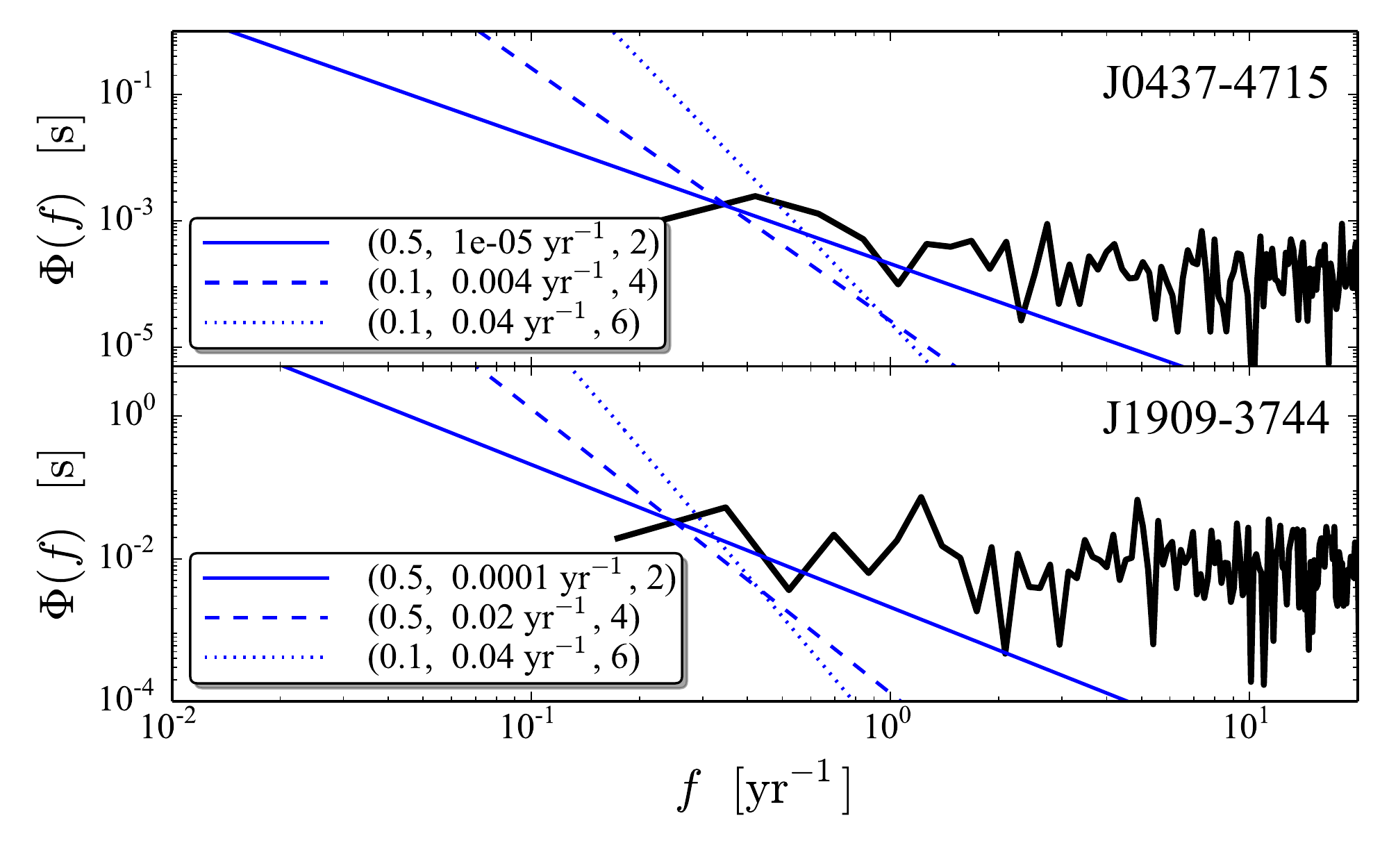}
	\caption{\label{superfluidPSD2} Same as for figure \ref{superfluidPSD1}, but for PSRs J0437-4715 and J1909-3744, two pulsars from the PPTA catalogue with the lowest level of timing noise.  Theoretical curves for the superfluid turbulence model require smaller values of $f_{c}\approx\eta(R^{-1})$ than in figure \ref{superfluidPSD1} to remain consistent with the data.
	}   
\end{figure}

In figure \ref{superfluidPSD2} we plot two further examples of millisecond pulsar phase residuals.  These objects exhibit the lowest level of timing noise in the PPTA sample.  For the superfluid turbulence model to remain consistent with these data, the objects must have long decorrelation time-scales, i.e., $f_{c}\lesssim10^{2}$ yr$^{-1}$.  The data show the white noise component, $A_{\rm W}$, and the turbulence-driven red-component sits below $A_{\rm W}$.  Under these circumstances, the turnover in $\Phi_{\rm TN}(f)$ occurs too low in frequency to be observed, and the main factor limiting PTA detection is $A_{\rm W}$.

\section{Optimal Pulsars}\label{optimal}
\begin{figure*}
	\includegraphics[angle=0,width=0.95\linewidth]{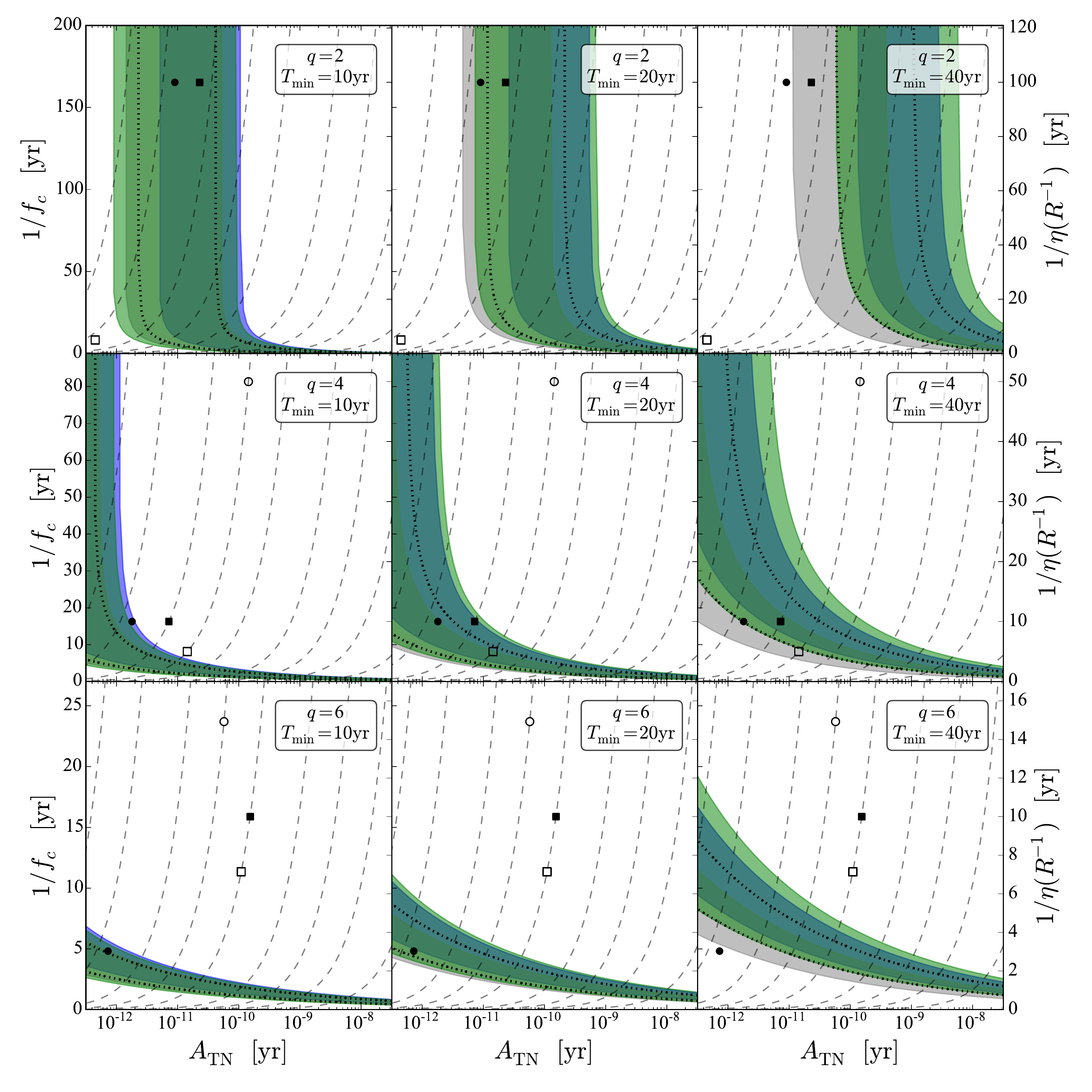}
	\caption{\label{fcAtn} Reciprocal of the timing noise turnover frequency, $1/f_{c}$, (left-hand vertical axis), as a function of the normalisation, $A_{\rm TN}$, (horizontal axis) for equation (\ref{phi}), for different values $T_{\rm min}$ as defined by equation (\ref{Tmin}) and $q$.  The right-hand vertical axis shows the decorrelation time-scale in the context of the superfluid turbulence model in section \ref{superfluidmodel}.  The dashed grey curves are $\lambda$ contours with $\lambda=1$, 0.5, 0.2, 0.1, 0.05, 0.02, 0.01, 0.005 and 0.002 going from left to right in each panel.  The shaded blue, green and grey regions, the thick blue curve and the dotted black curves represent the same theoretical limits for SMBHs and cosmic strings as in figure \ref{ObsTime}.  Overplotted are the superfluid turbulence model fits to the PPTA observational data presented in figure \ref{superfluidPSD1}, where the open circles, closed circles, open squares and closed squares are PSRs J1024$-$0719, J1643$-$1224, J1824$-$2452A and J1939$-$2134 respectively.  	
	}   
\end{figure*}

What pulsars are best placed to detect a gravitational wave background, given the longest time one is prepared to wait?  In figure \ref{fcAtn} we plot $1/f_{c}$ against $A_{\rm TN}$, for different values of $q$ and $T_{\rm min}$ in each panel.  The left-hand vertical axis displays the results for the model-independent form of $\Phi_{\rm TN}$ in equation (\ref{phi}).  The right-hand vertical axis registers the decorrelation time $1/\eta(R^{-1})$, in the superfluid turbulence model in section \ref{superfluidmodel}.  The dashed grey curves are curves of constant $\lambda$.  Overplotted are the superfluid turbulence model `fits' to the PPTA pulsar data in figure \ref{superfluidPSD1}, where the open circles, closed circles, open squares and closed squares are PSRs J1024$-$0719, J1643$-$1224, J1824$-$2452A and J1939$-$2134 respectively.

Figure \ref{fcAtn} allows us to ask whether, for example, $20$ yr of timing a specific pulsar will allow for sensitivity  to the most optimistic SMBH gravitational wave strain of $A_{\rm GW}=2.4\times10^{-15}$.  In the middle set of panels, the latter strain limit appears as the 
right-most boundary of the blue shaded region.  A pulsar with timing noise below this curve is sensitive to a gravitational wave signal in $T_{\rm obs}\le20\,{\rm yr}$.  Sensitivity depends on $q$ as illustrated in the three different panels running vertically.  It also depends on $f_{c}$.  For example, a hypothetical pulsar with $q=4$ and $A_{\rm TN}\approx10^{-11}$ yr is only sensitive to a gravitational wave background for $1/f_{c}\lesssim12$ yr.  This is an interesting constraint: a pulsar 
in a PTA that tolerates $T_{\rm min}\le20$ yr is sensitive to a gravitational wave background if $\Phi_{\rm TN}(f)$ exhibits a plateau after $\lesssim12$ yr of timing.

The superfluid turbulence model fits from figure \ref{superfluidPSD1} give an indication as to the usefulness of individual pulsars from the PPTA dataset.  For example, consider PSR J1939$-$2134 (closed squares).  If one again tolerates $T_{\rm min}\le20$ yr, the fits imply 
a pulsar is sensitive to a conservative prediction for the gravitational wave background
for $q\lesssim4$, although for $q\approx4$ this requires the timing noise spectrum to plateau after approximately 15 yr of timing.  We emphasise again that the model fits should only be taken as indicative; careful and detailed analysis is required to extract the true timing noise signal parameters from the data.

\section{Conclusion}\label{conclusion}
Pulsar Timing Array limits on the cosmological gravitational wave background are continually dropping to the point where they usefully constrain galaxy formation models \citep{shannon13a}.  
Positive detections, on the other hand, require a cross-correlation algorithm to simultaneously analyse timing residuals from multiple pulsars.  
Such a detection will likely occur when the gravitational wave background is the largest component in the unmodelled portion of many individual pulsar's timing residuals \citep{siemens13}.
If the timing noise spectrum is steeper asymptotically (at high $f$) than the gravitational wave spectrum, this is only possible if the timing noise spectrum flattens below some frequency, $f_c$.  
In this paper, we calculate the minimum observation time required, given $f_c$, before the gravitational wave background rises above the timing noise plateau in any specific pulsar.  We calculate this minimum observation time both in a model-independent way, and for timing noise arising from superfluid turbulence.  The latter model is selected not because it is necessarily preferred physically, but because it is simple, predictive and analytically tractable and therefore provides a test-bed for repeating the calculation with other physical models in the future.  

Our results rely on the timing noise spectrum whitening below some threshold frequency, $f_c$.  This provides an observational diagnostic that can be used to infer 
the effectiveness of an individual pulsar in a PTA.
If, upon observing a pulsar for some $T>1/f_c$, one finds that $\Phi_{\rm TN}(f)$ has not whitened below $f_c$, that 
pulsar's capacity for assisting usefully in the detection of a gravitational wave background is severely diminished.
The $f_c$ for a given pulsar is a function of the rotational parameters of the pulsar, and the gravitational wave amplitude and spectral index.  Therefore, using the prescription outlined in this paper, one can predict $f_c$ for a given pulsar and a given gravitational wave background.

In reality, measuring $f_{c}$ in a single pulsar is difficult.  Firstly, the noise in a given pulsar timing power spectrum is large, and secondly, the power in the lowest-frequency bin is generally dominated by the fact that a quadratic polynomial is fit to the timing residuals [see \citet{vanhaasteren13}].  These two effects potentially mimic a low-frequency turn-over, implying multiple low-frequency bins are required to confirm the existence of a low-frequency cut-off.  

Many data analysis algorithms simultaneously fit the timing model and the unknown noise contributions for any individual pulsar.  In this sense, one can include a low-frequency plateau into gravitational-wave detection algorithms, e.g., by way of a Bayesian prior on the form of the power spectral density.  Physically motivated models for timing noise, such as the superfluid turbulence model discussed herein, could be used to guide such priors.

\section*{acknowledgments}
We are grateful to the anonymous reviewer for the thoughtful and thorough review of the manuscript.
PDL and AM are supported by Australian Research Council (ARC) Discovery Project DP110103347.  PDL is also supported by ARC DP140102578.  VR is a recipient of a John Stocker Postgraduate Scholarship from the Science and Industry Endowment Fund.  We thank Yuri Levin for comments on the manuscript and Ryan Shannon for comments on an earlier version.  Calculations of the cosmic string stochastic background used the GWPlotter website: \hyperref[http://homepages.spa.umn.edu]{http://homepages.spa.umn.edu/$\sim$gwplotter}.

\bibliographystyle{mn2e}
\bsp
\label{lastpage}

\pagerange{\pageref{firstpage}--\pageref{lastpage}}
\maketitle



\end{document}